\documentclass[conference]{IEEEtran}

\ifCLASSINFOpdf
\usepackage[pdftex]{graphicx}
\usepackage{subcaption}
\usepackage{color}
\usepackage{aas_macros}
\usepackage{gensymb}
\usepackage{array}
\usepackage{lineno}
\usepackage{balance}
\usepackage{amsmath}
\usepackage{tikz}
  \DeclareGraphicsExtensions{.pdf,.jpeg,.png}
\else
\fi

\makeatletter
\newcommand{\thickhline}{%
    \noalign {\ifnum 0=`}\fi \hrule height 1pt
    \futurelet \reserved@a \@xhline
}
\newcolumntype{"}{@{\hskip\tabcolsep\vrule width 1pt\hskip\tabcolsep}}
\makeatother

\begin{document}

\title{Which is better, a SCoTSS gamma imager, or an ARDUO UAV-borne directional detector?}

\author{\IEEEauthorblockN{Andrew~McCann\IEEEauthorrefmark{1}, 
 Laurel~E.~Sinclair\IEEEauthorrefmark{1},
 Patrick~R.B.~Saull\IEEEauthorrefmark{2},
 Christian~Van~Ouellet\IEEEauthorrefmark{3},
 Richard~Fortin\IEEEauthorrefmark{1},
 Carolyn~Chen\IEEEauthorrefmark{1},\\
 Maurice~J.~Coyle\IEEEauthorrefmark{1},
 Rodger~Mantifel\IEEEauthorrefmark{2},
 Audrey~M.L.~MacLeod\IEEEauthorrefmark{2},
 Reid~A.~Van~Brabant\IEEEauthorrefmark{1},
 John~Buckle\IEEEauthorrefmark{1},\\
 Pierre-Luc~Drouin\IEEEauthorrefmark{3},
 Jens~Hovgaard\IEEEauthorrefmark{4},
 Bohdan~Krupskyy\IEEEauthorrefmark{4},
 Blake~Beckman\IEEEauthorrefmark{3}, and 
 Blaine~Fairbrother\IEEEauthorrefmark{3}}
\IEEEauthorblockA{\IEEEauthorrefmark{1} Canadian Hazard Information Service, Natural Resources Canada, Ottawa, Canada\\}
\IEEEauthorblockA{\IEEEauthorrefmark{2} Measurement Science and Standards portfolio of the National Research Council Canada, Ottawa, Canada\\}
\IEEEauthorblockA{\IEEEauthorrefmark{3} Defence Research and Development Canada, Ottawa, Canada\\}
\IEEEauthorblockA{\IEEEauthorrefmark{4} Radiation Solutions Inc., Mississauga, Canada}}

\maketitle

\begin{abstract}
The SiPM-based Compton Telescope for Safety and Security (SCoTSS) has
been developed with inorganic crystalline scintillator material for
gamma detection.  The instrument is sensitive enough to be used in a
mobile survey mode, accumulating energy deposited in any crystal
second-by-second and tagging these spectra with GPS position.  The
SCoTSS imager of course has the additional advantage of being able to
produce an image of the radioactive objects in its field of view using
events that satisfy a coincidence trigger between the scatter and
absorber layers.  The Advanced Radiation Detector for UAV Operations
(ARDUO) on the other hand, is a non-imaging directional detector
intended for use aboard a small unmanned aerial vehicle (UAV).  The
ARDUO detector features exactly the same volume of CsI(Tl) as is used
in the absorber layer of a single SCoTSS module, giving it similar
detection and alarming sensitivity, and map-making capability.
However, in the ARDUO detector, the crystals are arranged closely
together to optimize direction determination from self-shielding
effects.  Flown in a grid pattern with a UAV over an area of extended
contamination, the ARDUO detector is also capable of making a map or
image of that area.  With its close-packed crystal arrangement, the
ARDUO detector makes a poor Compton imager but does have some ability
to produce a peripheral image in a fly-by. In this presentation we
investigate the relative merits of Compton imaging versus mobile
directional detection.
\end{abstract}

\IEEEpeerreviewmaketitle

\section{Introduction}
The detection of ionizing radiation emitted by radioactive material
using instruments employing scintillating crystals and light sensors
has a long history \cite{2002JLum..100...35W}. Radionuclide
identification is typically performed by examining peaks found in the
measured energy spectrum and comparing them to known radionuclide
emission lines. For a stationary detector, the measurement of the
radiation directionality, and thus the localization of the radioactive
material, can be achieved through the use of passive absorbing
material (e.g. collimation, coded aperture, etc), active absorbing
material (e.g. self-shielding) or through kinematic interactions
(e.g. Compton scattering). Surveying with a non-stationary instrument
allows for localization by simply measuring the activity at a variety
of points over some spatial region. These localization methods, by
their very nature, have intrinsic strengths and weaknesses which make
them particularly well suited, or ill suited, to specific
applications.

We have developed two instruments based on thallium-doped cesium
iodide CsI(Tl) crystal scintillators coupled to silicon
photomultiplier (SiPM) light sensors. While both instruments employ
almost identical sensing and read-out components and have similar
mass, one is configured as a Compton imaging telescope and the other as
a non-imaging self-shielding directional detector. Access to both
these instruments affords us a unique opportunity to examine and
compare their performance in a variety of deployment settings. In
these proceedings we describe our findings based on our study of these
instruments.


\begin{figure}[t]
\centering
\begin{subfigure}{0.245\textwidth}
\centering
\begin{picture}(120,125)\put(-7,0){
\includegraphics[width=0.98\textwidth]{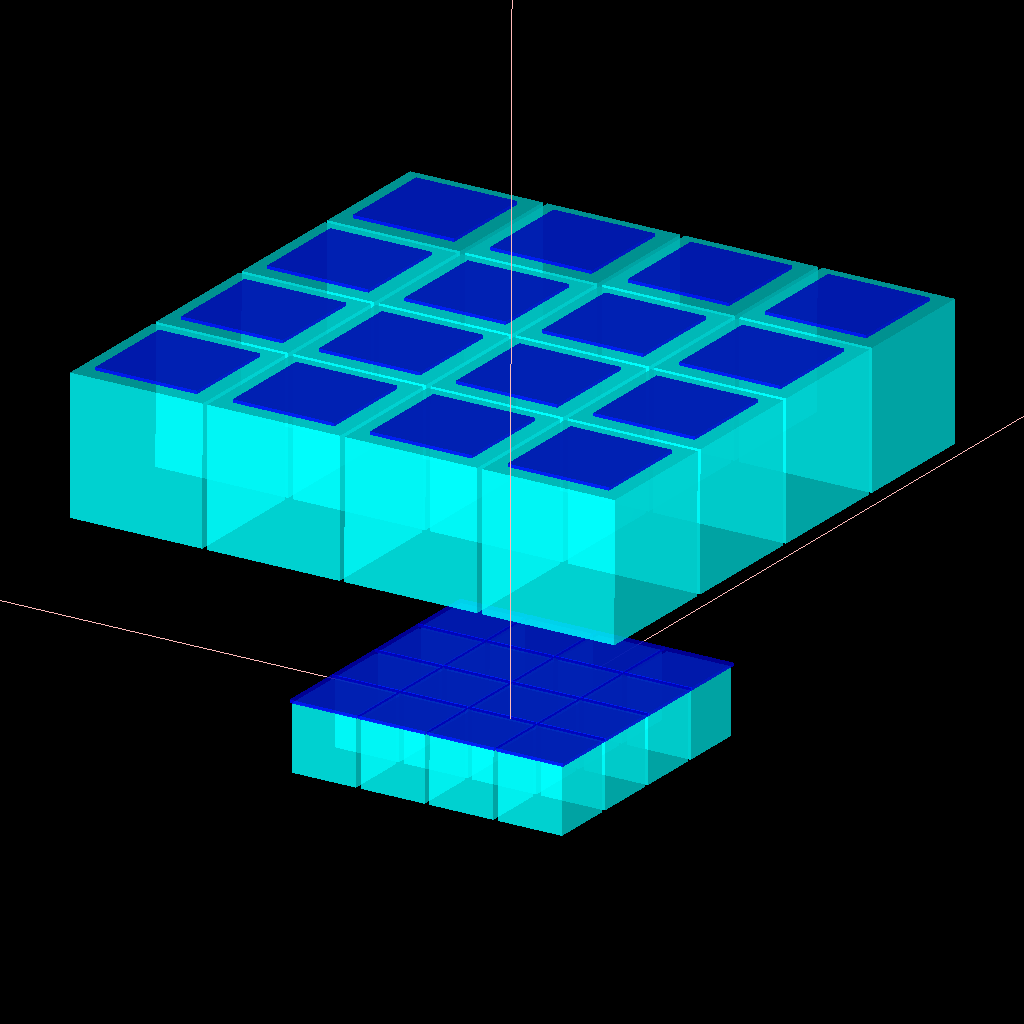}}\put(33,18){\textcolor{white}{\bf front}}\put(75,98){\textcolor{white}{\bf rear}}
\end{picture}
\caption{SCoTSS Compton imager}
\end{subfigure}%
\begin{subfigure}{0.245\textwidth}
\centering
\begin{picture}(120,125)\put(-7,0){
\includegraphics[width=0.98\textwidth]{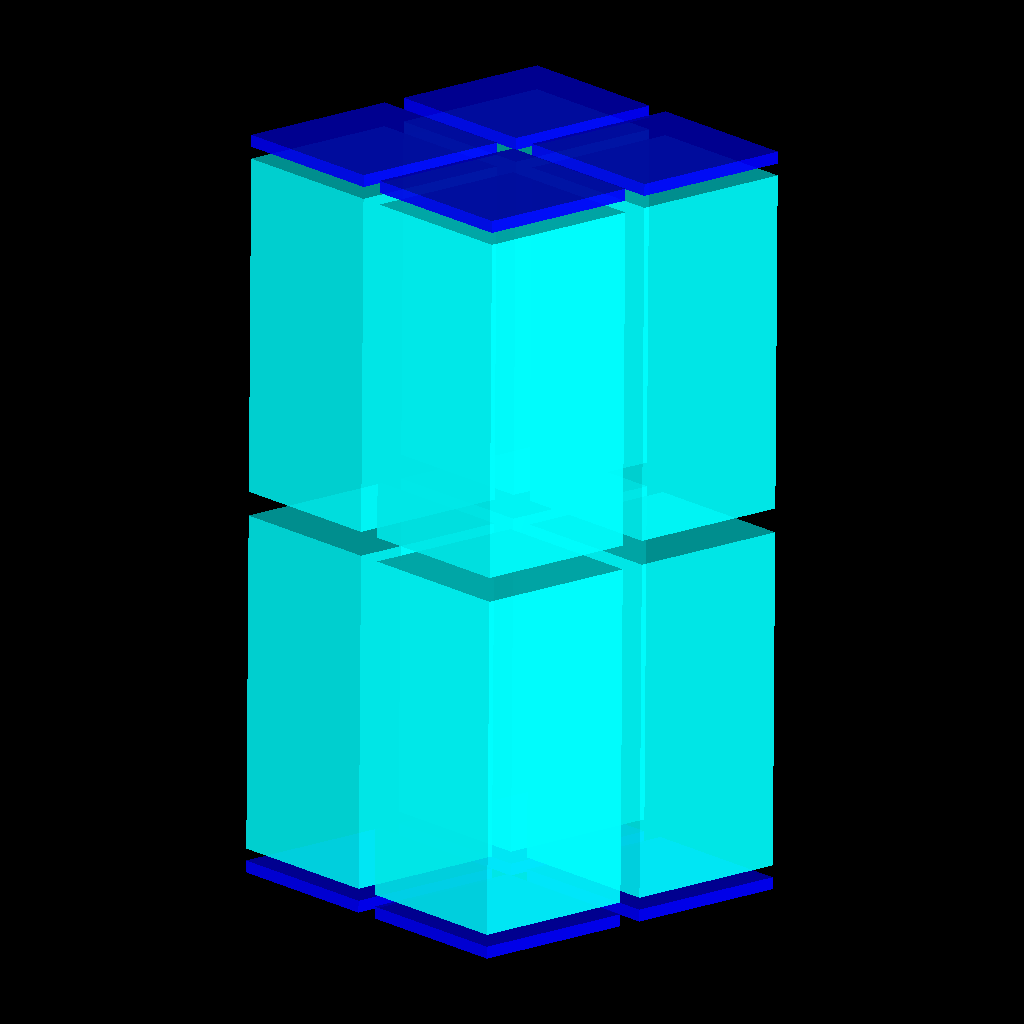}}\put(25,5){\textcolor{white}{\bf front}}\put(75,112){\textcolor{white}{\bf rear}}
\end{picture}
\caption{ARDUO
  directional detector}
\end{subfigure}
\caption{Schematic renderings of the active components of the two
  detectors. The SCoTSS Compton imager module is composed of two layers, each
  segmented into 4$\times$4 crystal arrays. The ARDUO directional detector is
  arranged as a close-packed segmented tower with two 2$\times$2 crystal
  arrays. To account for the different instrument sizes, image scales are
  unequal. The scintillator mass used in ARDUO is identical to
  the scintillator mass used in SCoTSS absorber (rear) layer.}
\label{fig_det}
\end{figure}
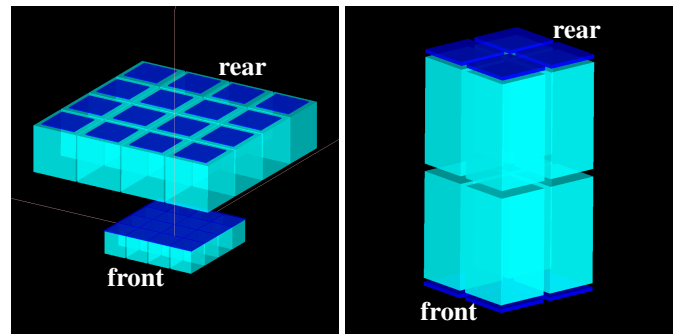

\section{Instruments}
\subsection{Compton Imager - SCoTSS}
The SiPM-based Compton Telescope for Safety and Security, SCoTSS, is a
Compton imaging radiation detector comprised of CsI(Tl) crystal
scintillators coupled to silicon photomultiplier light sensors
\cite{2014ITNS...61.2745S}. Simulation studies and prototyping
development over many years
\cite{2009ITNS...56.1262S,2010SPIE.7665E..1ES} have led us to a design
with two segmented layers. A ``scatter layer'', positioned at the
front of the detector, is designed to ensure Compton scatter
interactions yield excellent positional and energy resolution, while
minimizing the probability of multiple scatters within the front
layer. This layer is composed of cubic crystals, 1.35~cm on a side,
mated to SiPMs manufactured by SensL Technologies Ltd using optical
gel. The second layer, the ``absorber layer'', employs cubic crystals
2.8~cm on a side, also mated to SiPMs. This layer is designed to
absorb all of the remaining energy of gamma rays which scattered in
the front layer, while achieving the required level of positional
resolution. These two segmented sensor layers, connected with custom
designed coincidence-timing and read-out electronics, provide all the
information needed to use the Compton equation
\cite{1923PhRv...21..483C} to determine the incident angle of gamma
rays which scatter in the first layer and are fully absorbed in the
second.

The SCoTSS instrument is designed to be modular, with a single module
employing a 4$\times$4 array of crystals in the scatter layer and a
4$\times$4 array in the absorber layer (see
Figure~\ref{fig_det}a). This design allows for different
configurations of SCoTSS modules to be combined and arranged depending
on what is suitable and practical in a given deployment setting. In
this work, however, only a single SCoTSS module is considered.

\subsection{Self-shielding directional detector - ARDUO}
The Advanced Radiation Detector for Unmanned Aerial Vehicle
Operations, ARDUO, is a directional gamma-ray detector arranged as a
close-packed segmented tower with two 2$\times$2 CsI(Tl) scintillator
crystal arrays (see Figure~\ref{fig_det}b). The crystals are
2.8~cm $\times$ 2.8~cm $\times$ 5.6~cm in size and mated with the same
SiPM optical sensors used in the SCoTSS instrument. The direction to a
source is determined through a self-shielding method. Light is
collected in the scintillator crystals in the same way as is employed
in the SCoTSS module, however, energy deposits are accumulated into
one-second duration histograms and coincidence information is not
utilized. Using the relative rate of energy depositions in each
crystal a range of algorithms can be utilized to find the most likely
direction to a source of emission.

Both the SCoTSS and ARDUO detectors use power supplies and read-out
electronics custom built by our partners at Radiation Solutions
Inc. ARDUO has an active volume the same as that of the SCoTSS
absorber layer and their sensor components are read out with very
similar acquisition electronics. Their particular configuration is
therefore the principle difference between these two instruments. This
allows us to directly compare their performance using a range of
methods across a variety of deployment scenarios.

\begin{table}
\centering
  \begin{tabular}{  l " l | l  }
 Source Type     & ARDUO                    & SCoTSS  \\
                & Localization Methods      & Localization Methods \\ \thickhline 
  Point         & Response-function search  & Response-function search\\
                &                           & Compton-cone fit$^\dag$ \\ \hline
    Extended    & Rate~survey              & Rate~survey \\ 
                & Simple direction finding & Back projection$^\dag$\\ 
                & Response-function fit     & Response-function fit\\
  \end{tabular}
\caption{Summary of the localization and directional methods
    used in this study. The $\dag$ symbol indicates methods which
    require coincidence timing.}
\label{tab_meth}
\end{table}

\begin{figure*}
\centering
\textbf{Point-source results}\par\medskip
\begin{subfigure}{0.33\textwidth}
\centering
\includegraphics[width=0.98\textwidth]{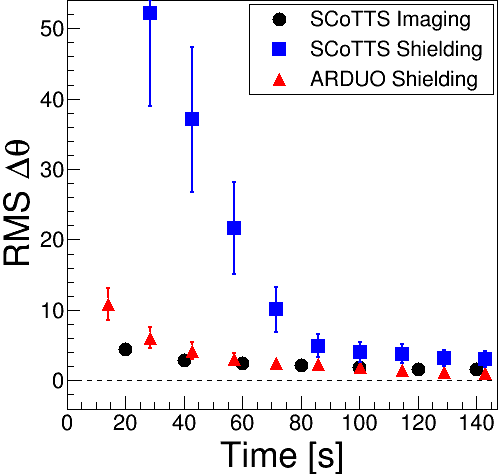} \caption{RMS deviation against time [10$^{\circ}$ offset]}
\end{subfigure}
\begin{subfigure}{0.33\textwidth}
\centering 
\includegraphics[width=0.98\textwidth]{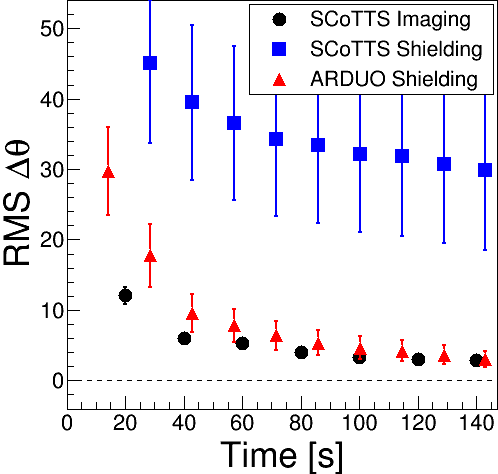} \caption{RMS deviation against time [60$^{\circ}$ offset]}
\end{subfigure}%
\begin{subfigure}{0.33\textwidth}
\centering
\includegraphics[width=0.98\textwidth]{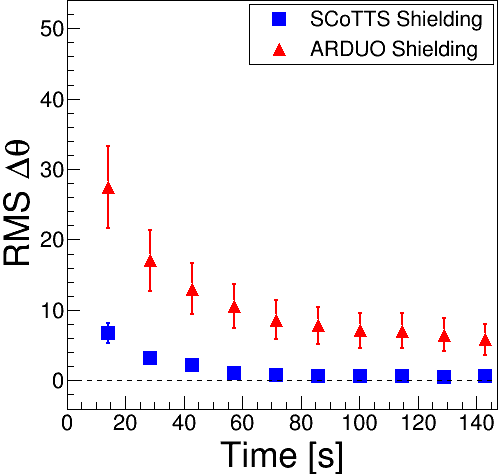} \caption{RMS deviation against time [90$^{\circ}$ offset]}
\end{subfigure}
\begin{subfigure}{0.33\textwidth}
\centering
\includegraphics[width=0.98\textwidth]{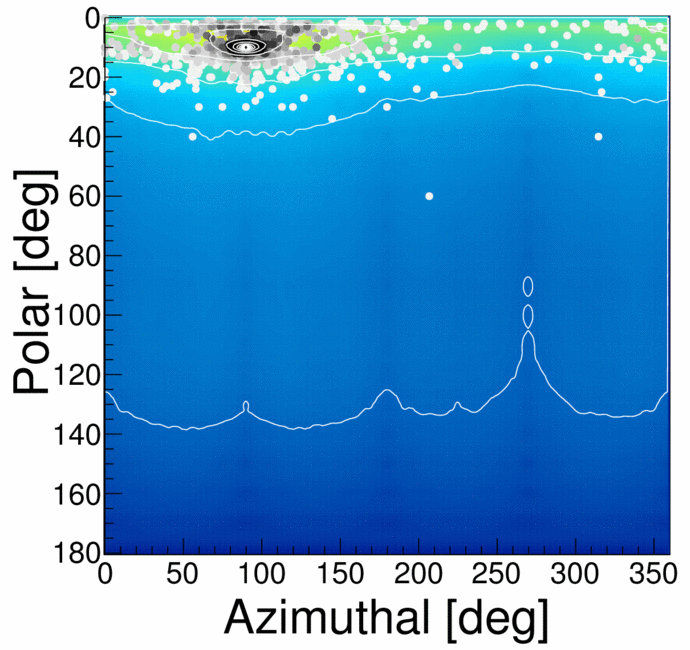} \caption{ARDUO likelihood [10$^{\circ}$ offset]}
\end{subfigure}%
\begin{subfigure}{0.33\textwidth}
\centering 
\includegraphics[width=0.98\textwidth]{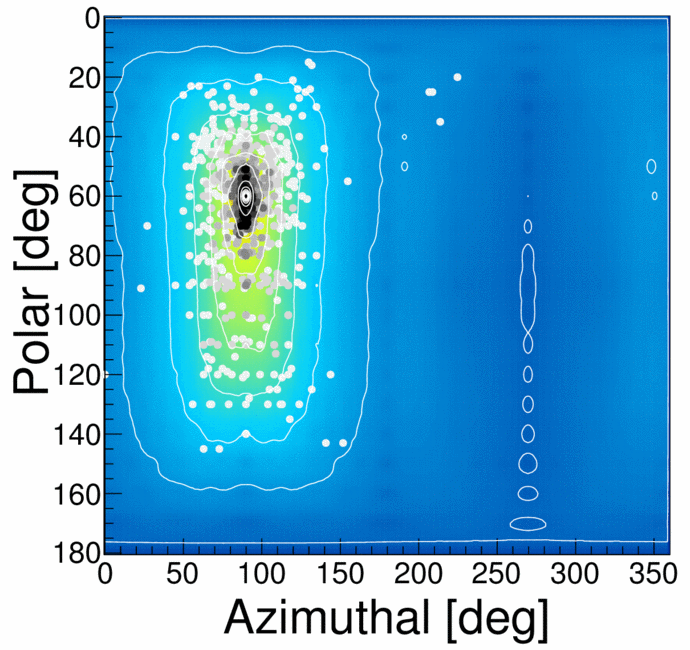} \caption{ARDUO likelihood [60$^{\circ}$ offset]}
\end{subfigure}
\begin{subfigure}{0.33\textwidth}
\centering
\includegraphics[width=0.98\textwidth]{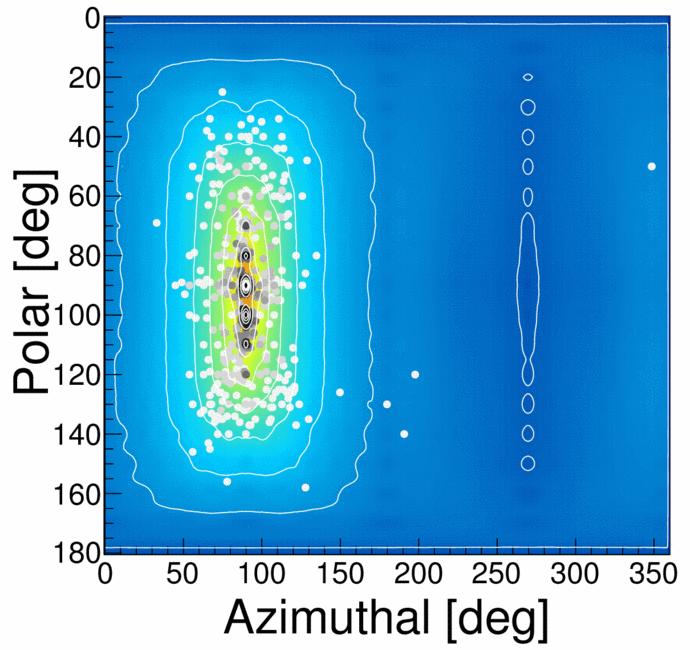} \caption{ARDUO likelihood [90$^{\circ}$ offset]}
\end{subfigure}
\begin{subfigure}{0.33\textwidth}
\centering
\includegraphics[width=0.98\textwidth]{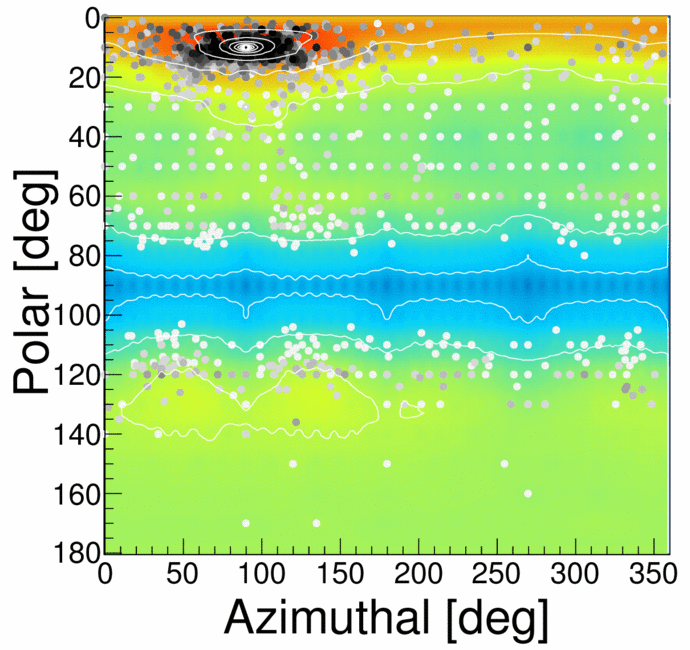} \caption{SCoTSS likelihood
  [10$^{\circ}$ offset]}
\end{subfigure}%
\begin{subfigure}{0.33\textwidth}
\centering 
\includegraphics[width=0.98\textwidth]{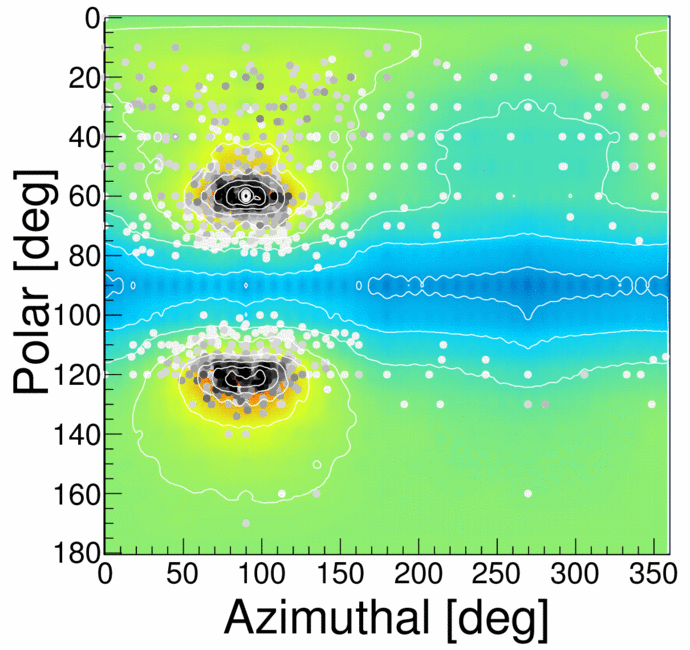} \caption{SCoTSS likelihood
  [60$^{\circ}$ offset]}
\end{subfigure}
\begin{subfigure}{0.33\textwidth}
\centering
\includegraphics[width=0.98\textwidth]{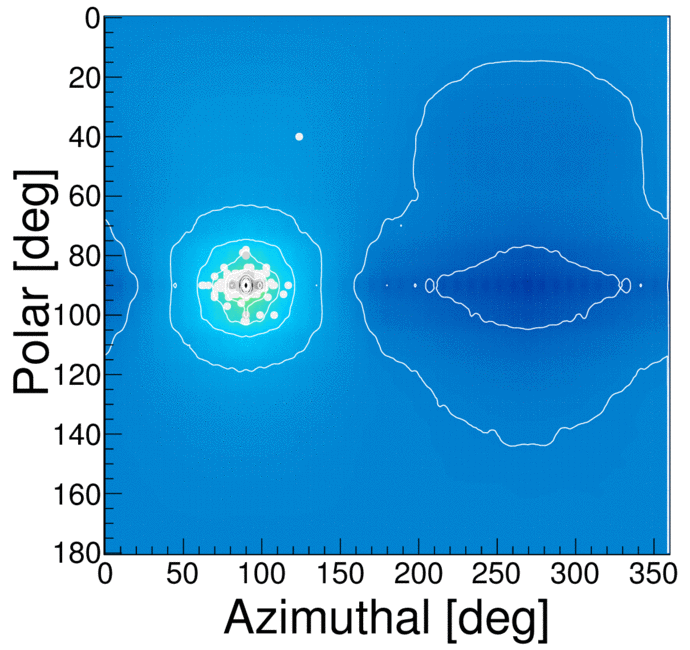} \caption{SCoTSS likelihood
  [90$^{\circ}$ offset]}
\end{subfigure}
\caption{Reconstruction of a simulated 1~mCi $^{137}$Cs point source
  located at an offset distance of 10~m using self-shielding and
  Compton imaging methods. Panels (a-c) show the root-mean-squared
  (RMS) deviation of the reconstructed direction to a point source
  versus accumulation time. The \textit{response-function search}
  self-shielding method (triangles and squares) can be compared to the
  \textit{Compton-cone fit} method (circles). The circular symbols in
  panels (d-f) show individual ARDUO self-shielding reconstructions,
  with their colour (from light-grey to black) indicating the
  acquisition time.  The colour map indicates the likelihood that the
  self-shielding algorithm will reconstruct a given position in 4$\pi$
  phase space when the true position is located at an azimuth of
  90$^{\circ}$ and a polar angle given in the individual sub-caption
  (see text Section~\ref{sss1}). Panels (g-i) show the equivalent
  information for the SCoTSS self-shielding reconstructions. Notable
  is the likelihood information in panel (h), indicating that the
  crystal configuration in the SCoTSS instrument yields areas of the
  response function look-up-table which are self-similar, leading to
  erroneous reconstructions which do not significantly improve with
  accumulation time. This is directly responsible for the large RMS of
  the SCoTSS self-shielding reconstruction plotted in panel (b). }
\label{fig_tti}
\end{figure*}
\section{Experimental Setup}
To examine and compare the performance of the SCoTSS and ARDUO
detectors we perform a range of simulation studies using the Geant4
\cite{2003NIMPA.506..250A} and EGSnrc simulation packages
\cite{EGSnrc1,EGSnrc2}. Both detectors are simulated in the presence
of point-like and distributed sources of photons emitted at 662~keV -
simulating the response to gamma-ray emission from $^{137}$Cs. Data
acquired with SCoTSS and ARDUO exposed to calibrated $^{137}$Cs
sources are used to verify our simulations, but not shown in this
work.

\subsection{Point source}
The point-like source studies examine the detector response to a 1~mCi
$^{137}$Cs source located at an offset distance of 10~m and positioned
at a range of angles from the principle detector axes.


\subsection{Distributed source}
We simulated 150~m $\times$ 150~m aerial grid surveys of distributed
sources located 10~m below the detectors.  Grid points are separated
by 10~m in both easting and northing corresponding to a flight speed
of 10~m/s and line spacing of 10~m. Two distributed-source sizes were
studied, one at the 100~m size scale consisting of an L-shaped source
with a 10~m $\times$ 60~m northing section and a 70~m $\times$ 5~m
easting section (see Figures~\ref{fig_L}a, c and e). The second
distributed source, 10 times smaller than the first, consists of an
L-shaped source with a 1~m $\times$ 6~m northing section and a 7~m
$\times$ 0.5~m easting section (see Figures~\ref{fig_L}b, d and f).

Table~\ref{tab_meth} summarizes the range of directional and imaging methods
investigated in this work.

\section{Method}
\subsection{Non-directional method - Rate survey}
This method does not use the individual rate information of each detector
crystal, but simply the total rate of energy deposits, and plots this total
rate against the location in the grid survey. Results based on this method are
used in the distributed-source studies plotted in Figure~\ref{fig_L}a and
Figure~\ref{fig_L}b.

\subsection{Self-shielding directional methods}
Self-shielding directional methods use the rate of energy deposits in each
scintillator crystal to determine the most likely direction to the source of
emission. One can imagine a simple case of two crystals side by side - crystal
\textit{A} on the left of crystal \textit{B}. If, in the presence of a source,
the rate in crystal \textit{A} is much higher than that in \textit{B} then one
can infer that a source is located somewhere to the left of \textit{B}, since
\textit{A} is shielding \textit{B} from the flux of incident events. A range
of methods based on the (relative) rate of energy deposits are used in this
study.

\subsubsection{Response-function search}\label{sss1}
Simulation datasets are computed with point sources located at a range
of angular positions covering the 4$\pi$ angular phase space around
the detector at some offset distance. Interpolating this dataset one
can construct look-up tables (LUT) of the detector response,
$S_{i}(\theta,\phi)$, indicating the expected relative rate in crystal
$i$ in the presence of a point source positioned at polar coordinate
angles $(\theta,\phi)$.  To localize a source at some unknown
position, the LUTs are searched to find the $(\theta,\phi)$-point
where the detector response is most similar to the measured rates.

The similarities between the detector response at one
$(\theta,\phi)$-point compared to another, $(\theta',\phi')$, is
indicative of the likelihood that this algorithm will reconstruct the
source location to be at $(\theta',\phi')$, when the true position is
at the location $(\theta,\phi)$. As a proxy for this likelihood, we
compute the $N$-dimensional distance in LUT rate-space between
$(\theta,\phi)$ and $(\theta',\phi')$,
\[
D(\theta',\phi') = \sqrt{\sum_{i=1}^{N}(S_{i}(\theta,\phi) - S_{i}(\theta',\phi'))^{2}}
\]
where $N$ is the number of crystals. This quantity allows us to assess the
self-shielding directional characteristics of a given detector configuration
from the LUT. For examples of the use of this distance metric see the colour
map in Figures~\ref{fig_tti}d-i.

\subsubsection{Simple direction finding}
One starts by choosing three perpendicular axes in the detector geometry,
e.g. ``top-bottom'', ``left-right'' and ``back-front''. The difference in the
measured rate projected along these directions is then computed, resulting in
a three-dimensional vector. If constructed sensibly, this vector provides a
crude estimate of the direction of the source of emission. It may not,
however, always be possible to choose perpendicular axes without biasing the
rate along some projected directions.

\subsubsection{Response-function fit}
In simulation, the rate of energy-deposition events in each crystal
is measured in response to 10~m $\times$ 10~m planer sources located at
a grid of locations offset from the detector.  With the appropriate
grid spacing and altitude chosen, a library of detector responses
suited to our particular deployment setting is thus compiled. The rate
of energy-deposition events measured in each crystal during an aerial
survey above some unknown distributed source is nothing more than the
true activity concentration folded with the (known) detector response.
Approximating the unknown distributed source as a tiled array of
individual 10~m $\times$ 10~m planer sources with floating activities,
and using the $\chi^2$-fit described in \cite{2015IEETRAN7581989S}, we
can measure the surface activity deconvolved with the detector
response.

The implementation of this method, as described here, has been chosen
to suit our particular applications - 150~m $\times$ 150~m aerial
surveys of distributed sources with 10~m line spacing and altitude.
We note that our implementation has no ability to resolve spacial
features below the 10~m $\times$ 10~m resolution used in constructing
the planer response library.

\subsection{Compton imaging methods}
Compton imaging methods require coincidence timing electronics and can
only reconstruct the directional information of a subset ($\sim$0.5\%)
of impinging events - those which scatter in the first layer and are
fully absorbed in the second. Though this requirement hugely reduces
the number of considered events, individual events which meet this
criteria contain a high degree of directional information. The Compton
equation allows for the reconstruction of the incident direction up to
an azimuthal degeneracy around the axis connecting the interaction
points in the scatter and absorber layers. Thus, the true incident
direction of each event is known to lie on the surface of a cone
projected outward from the detector along the interaction axis. In the
case of a single point source with zero background, only three events
are needed to uniquely define the location of the source within
measurement uncertainty.

\subsubsection{Back projection}
Back-projection methods start with the choice of an image plane in front of
the detector. The Compton cone projected outward from the detector will
intersect this plane and along the intersection points an elliptical annulus is
drawn. Successive annuli overlayed form an image of the emission.

\subsubsection{Compton-Cone Fit}
Assuming a common source of origin a $\chi^2$-minimization algorithm can be
applied to the back-projected Compton cones to find the common point of
closest approach \cite{2009ITNS...56.1262S}. This method allows for rigorous
treatment of the measurement uncertainty on the individual cones in the
reconstruction, permitting a source position measurement with high precision
and accuracy.
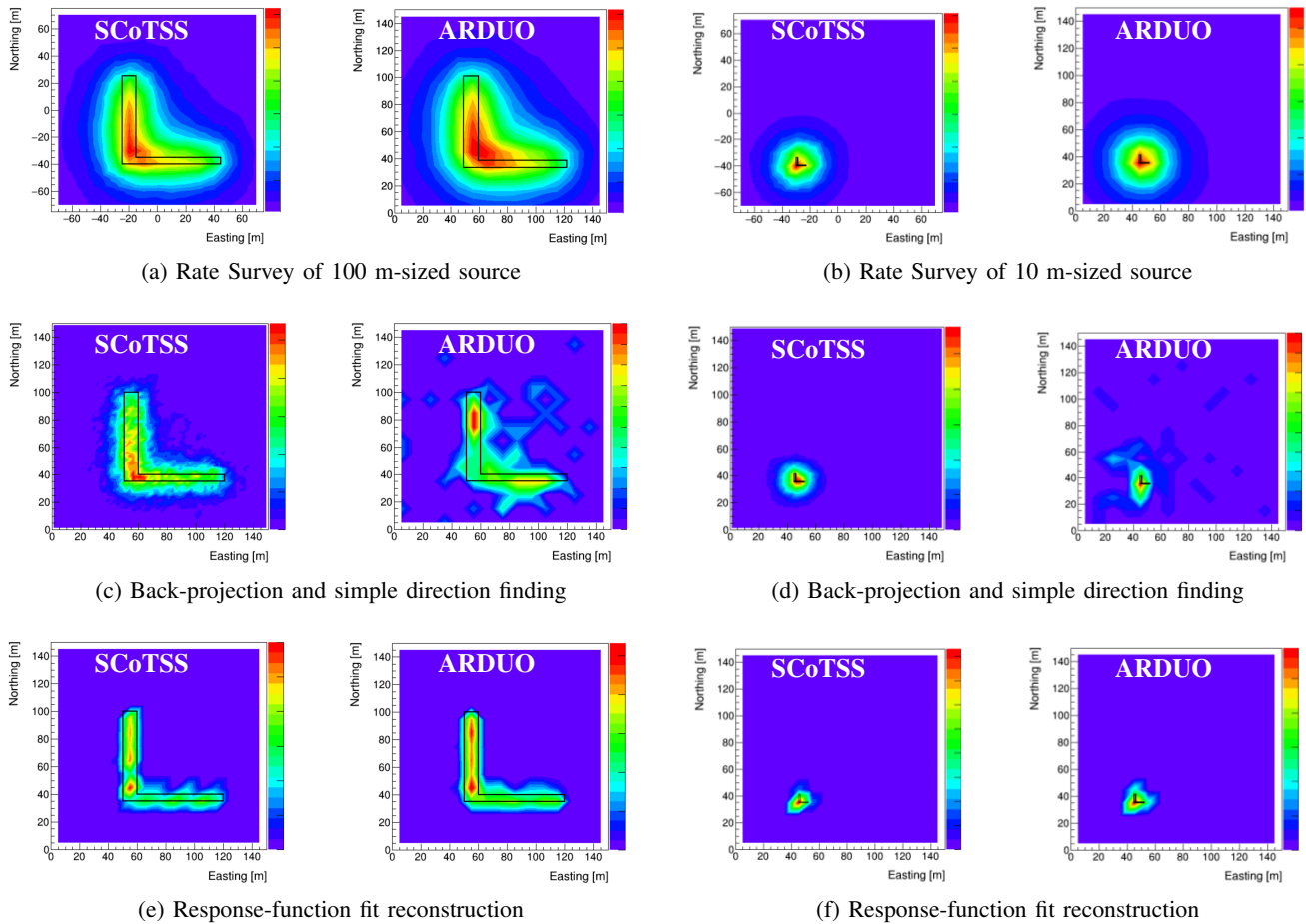
\begin{figure*}
\centering
\textbf{Distributed-source results}\par\medskip
\begin{subfigure}{0.49\textwidth}
\begin{picture}(220,105)
\put(1,0)
{
\begin{tikzpicture}
    \node[anchor=south west,inner sep=0] at (0,0) {\includegraphics[width=0.47\textwidth]{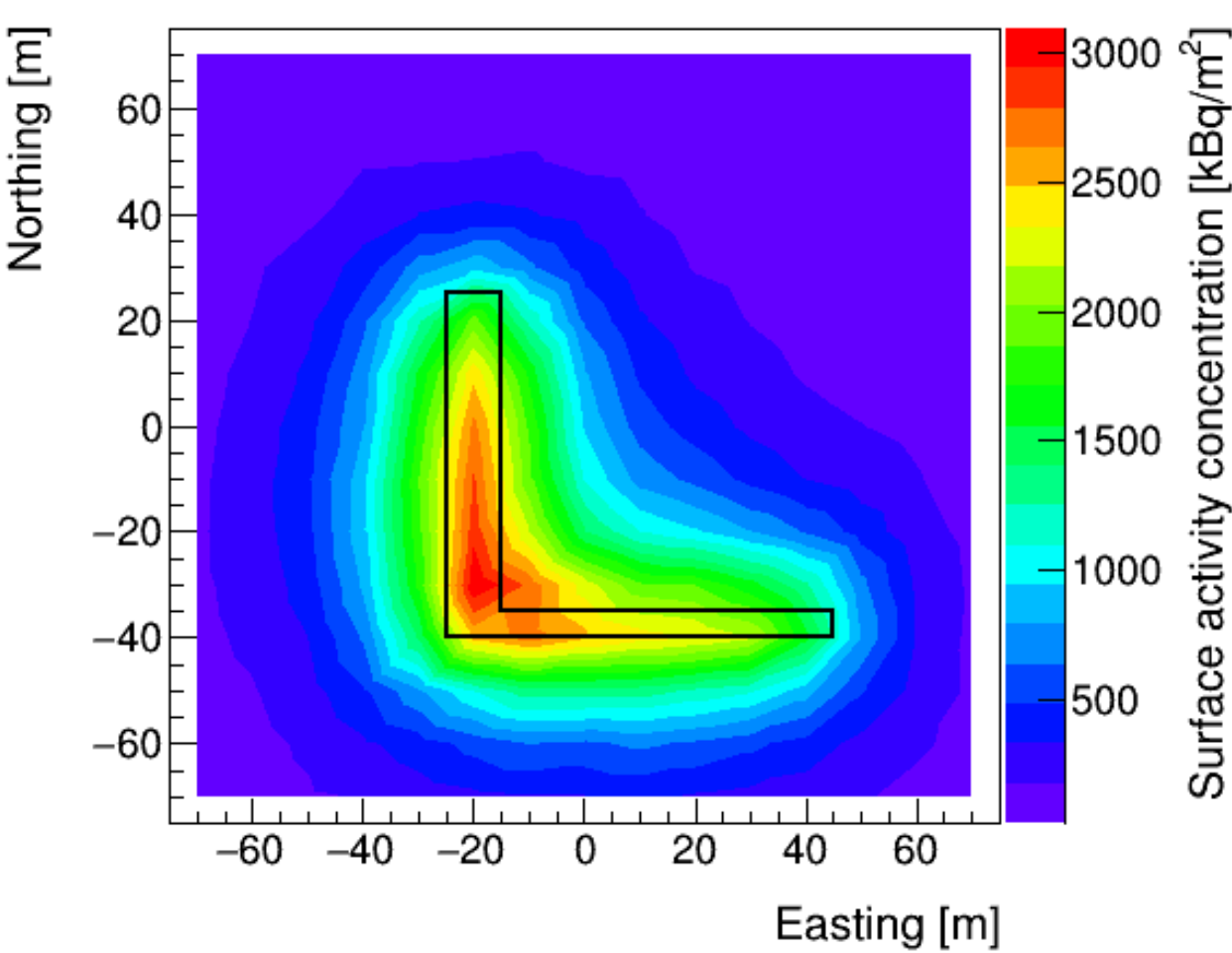}};
    \draw [fill=white,white] (3.64,0.1) rectangle (4.2,3.4); 
\end{tikzpicture}
}
\put(37,78){\textcolor{white}{\bf SCoTSS}}
\put(130,0)
{
\begin{tikzpicture}
    \node[anchor=south west,inner sep=0] at (0,0) {\includegraphics[width=0.47\textwidth]{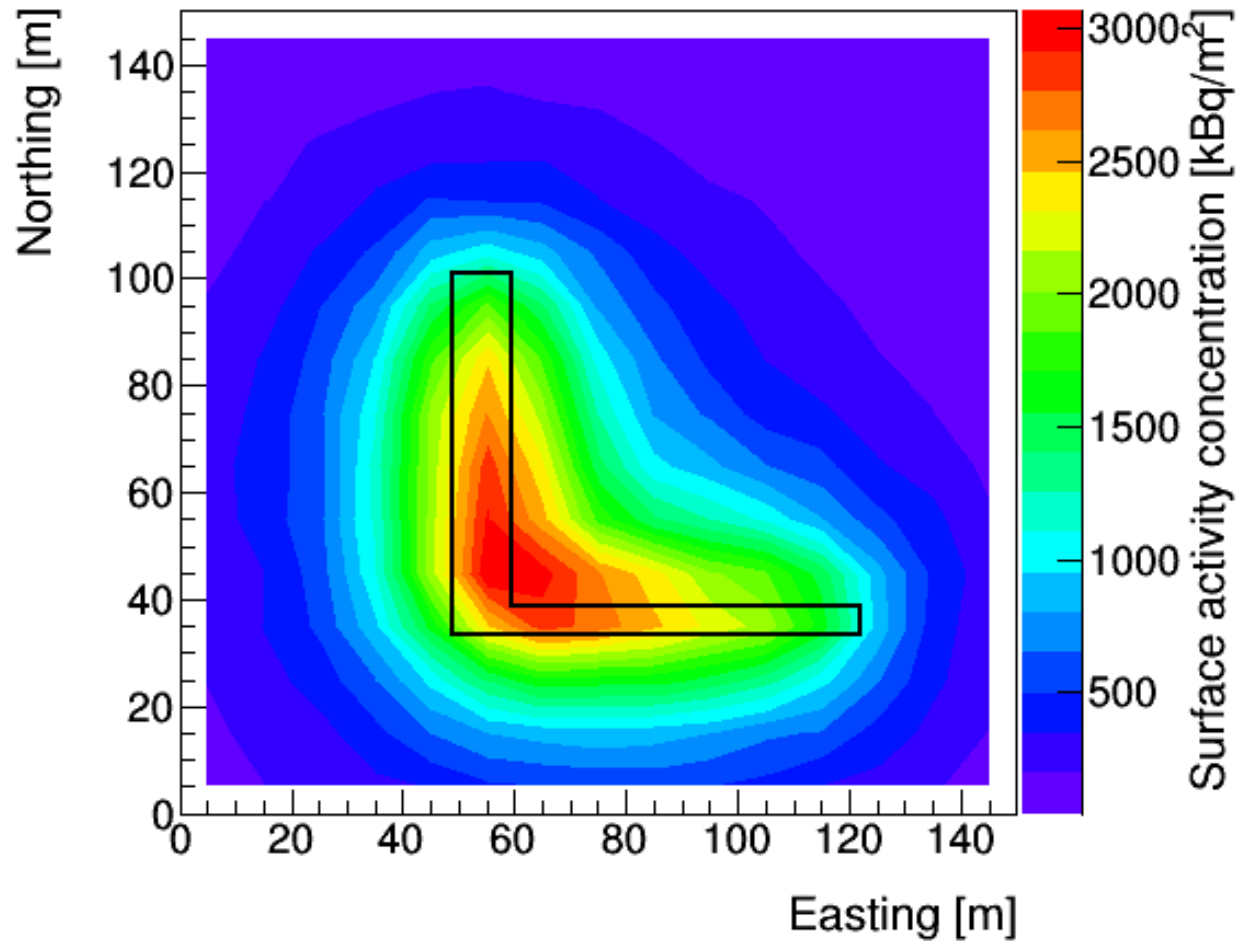}};
    \draw [fill=white,white] (3.67,0.1) rectangle (4.2,3.4); 
\end{tikzpicture}
}
\put(167,78){\textcolor{white}{\bf ARDUO}}\end{picture}
\caption{Rate Survey of 100~m-sized source}
\end{subfigure}
\begin{subfigure}{0.49\textwidth}
\begin{picture}(220,105)
\put(1,0)
{
\begin{tikzpicture}
    \node[anchor=south west,inner sep=0] at (0,0) {\includegraphics[width=0.47\textwidth]{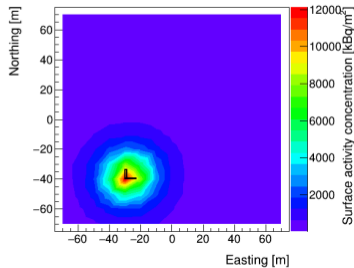}};
    \draw [fill=white,white] (3.64,0.1) rectangle (4.2,3.4); 
\end{tikzpicture}
}
\put(37,78){\textcolor{white}{\bf SCoTSS}}
\put(130,0)
{
\begin{tikzpicture}
    \node[anchor=south west,inner sep=0] at (0,0) {\includegraphics[width=0.47\textwidth]{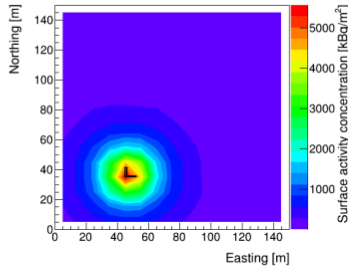}};
    \draw [fill=white,white] (3.7,0.1) rectangle (4.2,3.4); 
\end{tikzpicture}
}
\put(167,78){\textcolor{white}{\bf ARDUO}}\end{picture}
\caption{Rate Survey of 10~m-sized source}
\end{subfigure}
\begin{subfigure}{0.49\textwidth}
\begin{picture}(220,105)
\put(1,0)
{
\begin{tikzpicture}
    \node[anchor=south west,inner sep=0] at (0,0) {\includegraphics[width=0.47\textwidth]{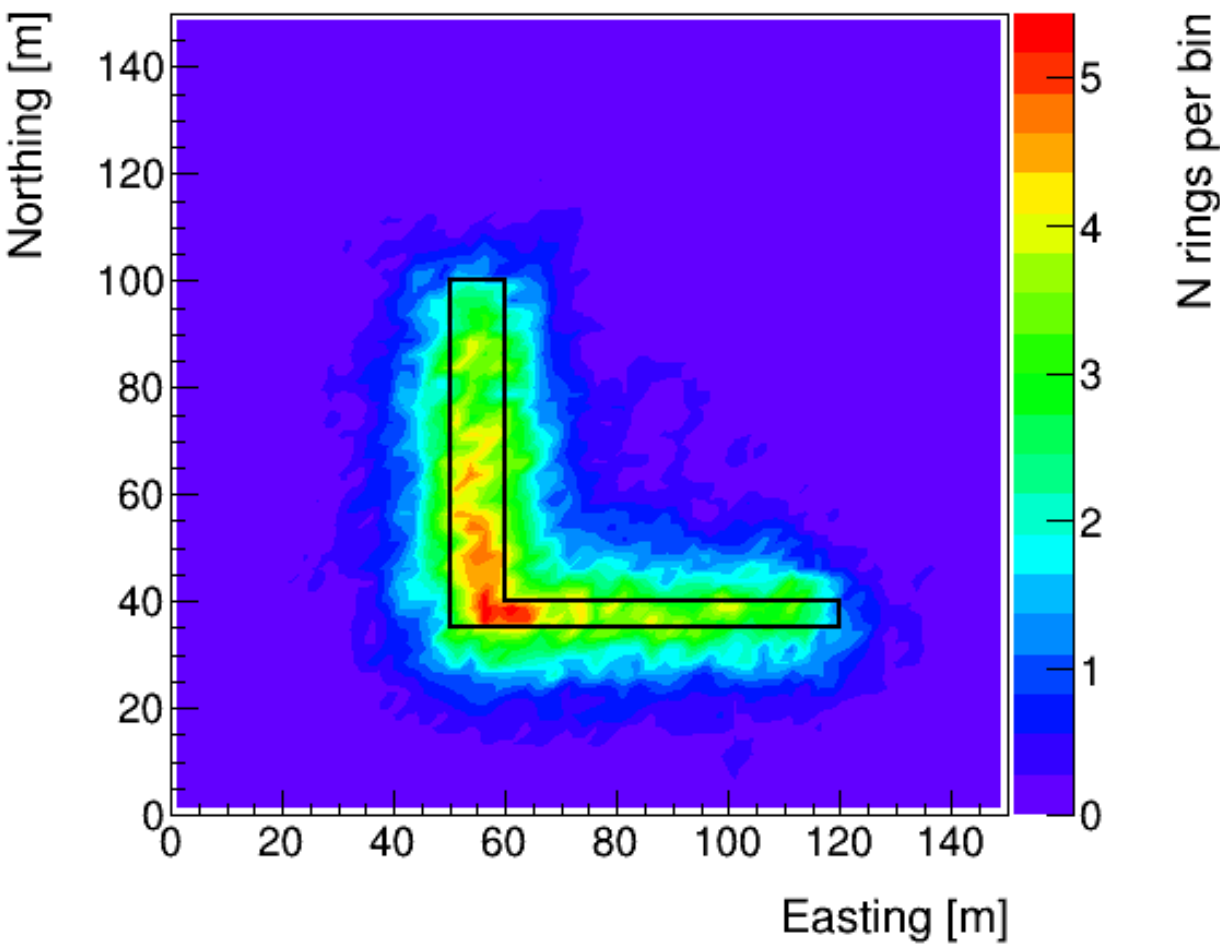}};
    \draw [fill=white,white] (3.7,0.1) rectangle (4.2,3.4); 
\end{tikzpicture}
}
\put(37,80){\textcolor{white}{\bf SCoTSS}}
\put(130,0)
{
\begin{tikzpicture}
    \node[anchor=south west,inner sep=0] at (0,0) {\includegraphics[width=0.47\textwidth]{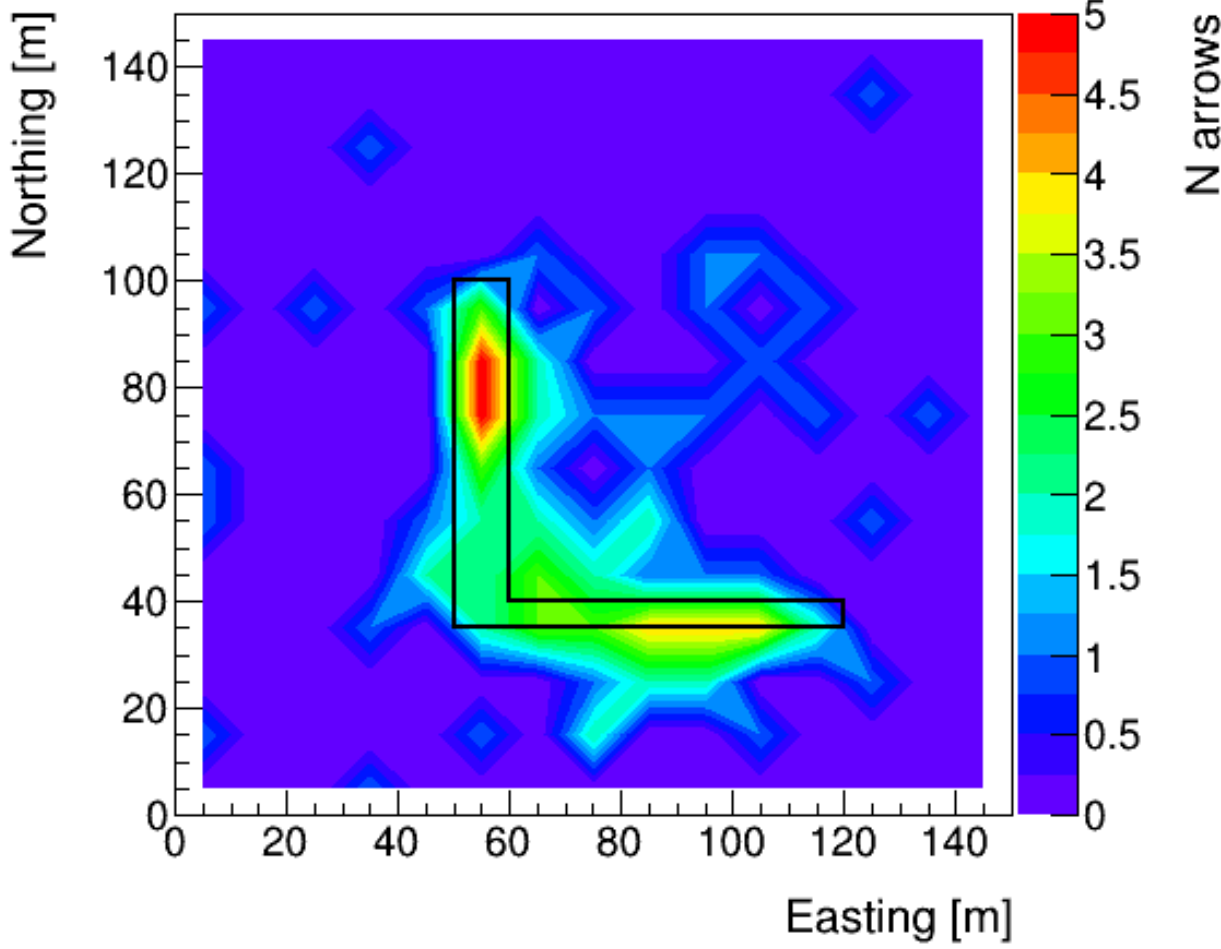}};
    \draw [fill=white,white] (3.7,0.1) rectangle (4.2,3.4); 
\end{tikzpicture}
}
\put(167,80){\textcolor{white}{\bf ARDUO}}\end{picture}
\caption{Back-projection and simple direction finding}
\end{subfigure}
\begin{subfigure}{0.49\textwidth}
\begin{picture}(220,105)
\put(1,0)
{
\begin{tikzpicture}
    \node[anchor=south west,inner sep=0] at (0,0) {\includegraphics[width=0.47\textwidth]{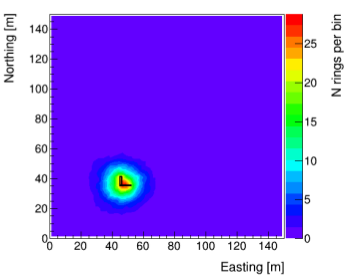}};
    \draw [fill=white,white] (3.67,0.1) rectangle (4.2,3.4); 
\end{tikzpicture}
}
\put(37,78){\textcolor{white}{\bf SCoTSS}}
\put(130,0)
{
\begin{tikzpicture}
    \node[anchor=south west,inner sep=0] at (0,0) {\includegraphics[width=0.47\textwidth]{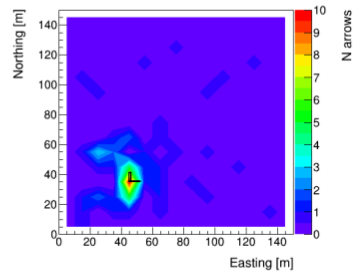}};
    \draw [fill=white,white] (3.68,0.1) rectangle (4.2,3.4); 
\end{tikzpicture}
}
\put(167,78){\textcolor{white}{\bf ARDUO}}\end{picture}
\caption{Back-projection and simple direction finding}
\end{subfigure}
\begin{subfigure}{0.49\textwidth}
\begin{picture}(220,105)
\put(1,0)
{
\begin{tikzpicture}
    \node[anchor=south west,inner sep=0] at (0,0) {\includegraphics[width=0.47\textwidth]{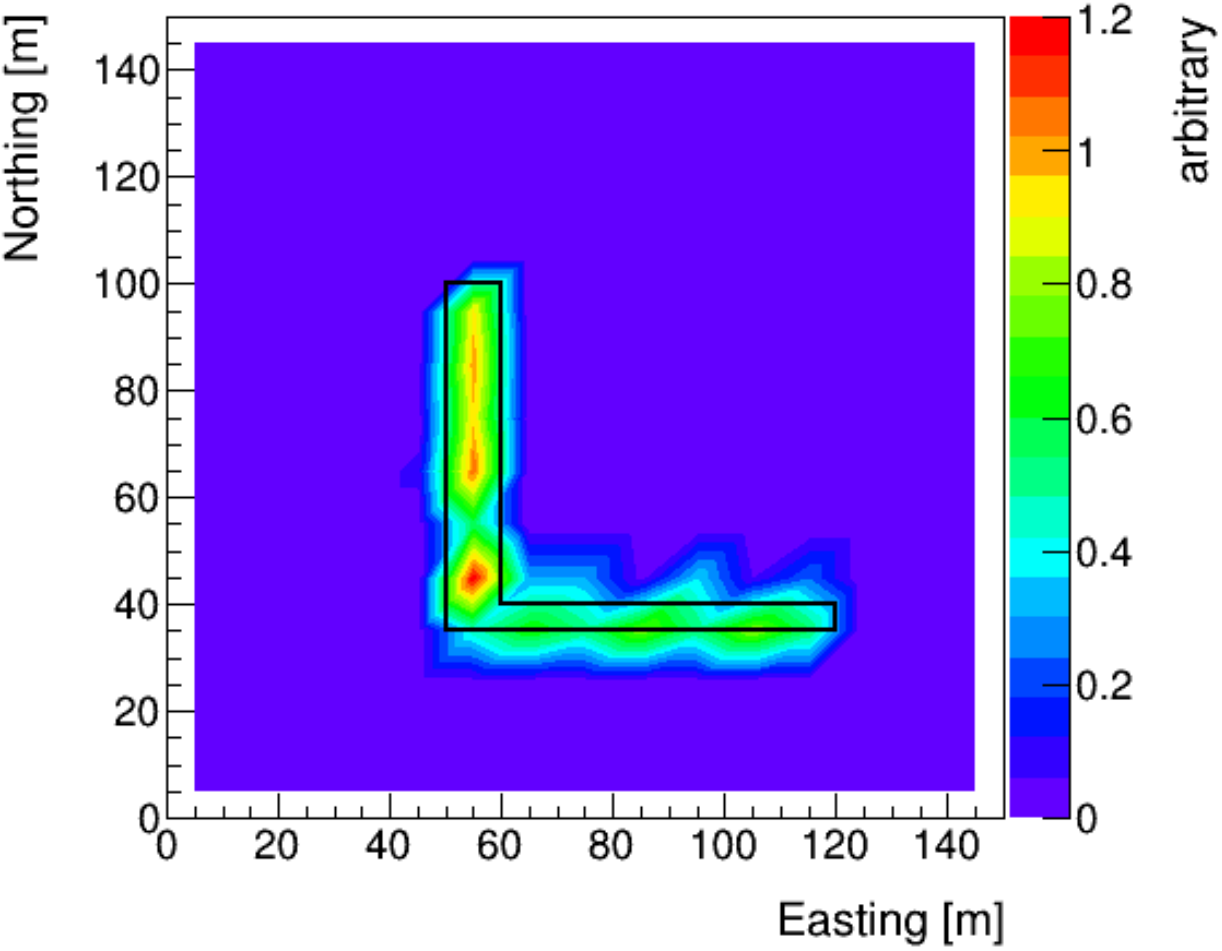}};
    \draw [fill=white,white] (3.68,0.1) rectangle (4.2,3.4); 
\end{tikzpicture}
}
\put(37,80){\textcolor{white}{\bf SCoTSS}}
\put(130,0)
{
\begin{tikzpicture}
    \node[anchor=south west,inner sep=0] at (0,0) {\includegraphics[width=0.47\textwidth]{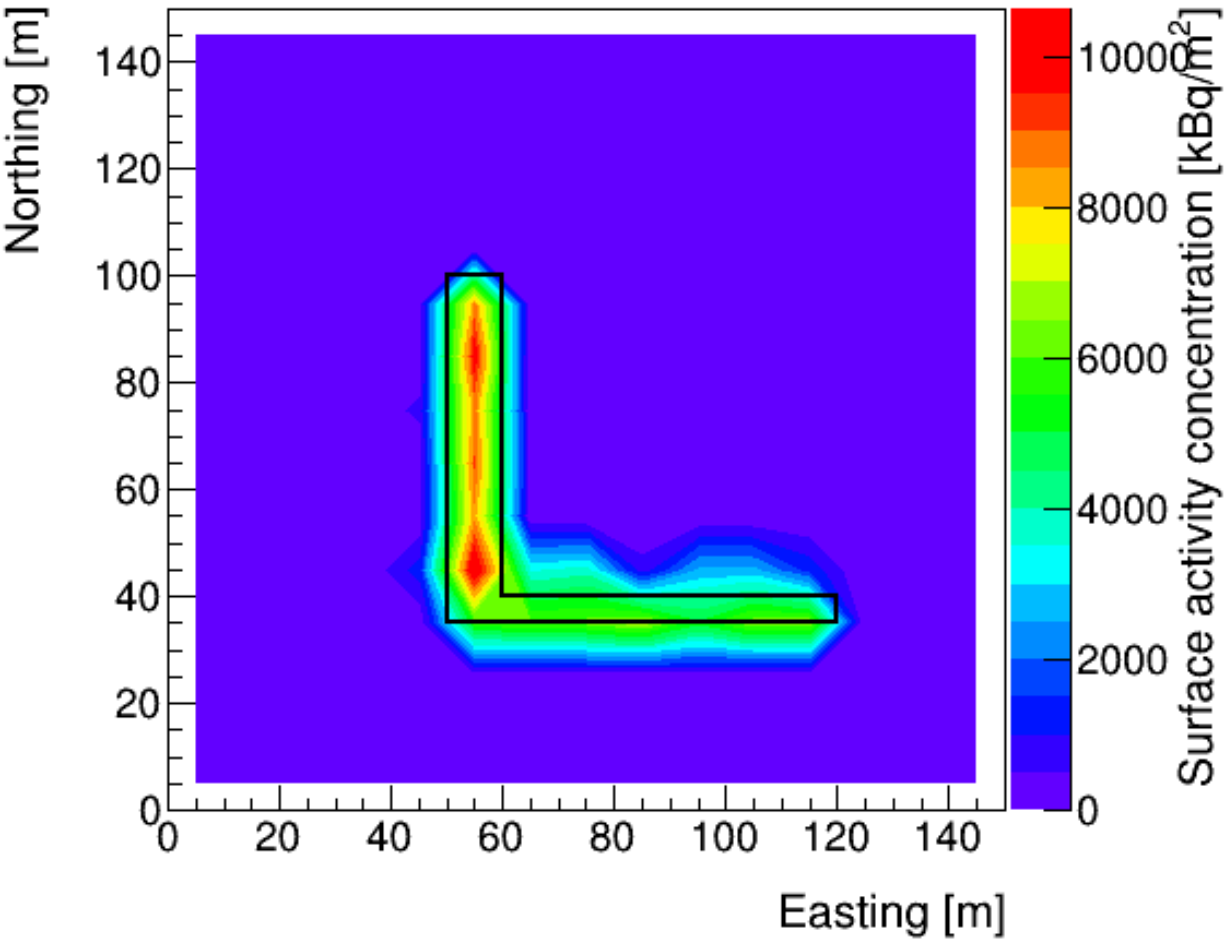}};
    \draw [fill=white,white] (3.685,0.1) rectangle (4.2,3.4); 
\end{tikzpicture}
}
\put(167,80){\textcolor{white}{\bf ARDUO}}\end{picture}
\caption{Response-function fit reconstruction}
\end{subfigure}
\begin{subfigure}{0.49\textwidth}
\begin{picture}(220,105)
\put(1,0)
{
\begin{tikzpicture}
    \node[anchor=south west,inner sep=0] at (0,0) {\includegraphics[width=0.47\textwidth]{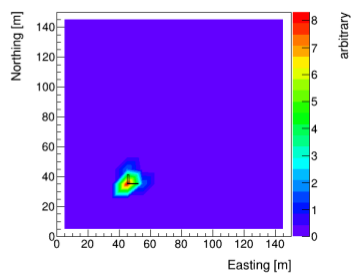}};
    \draw [fill=white,white] (3.68,0.1) rectangle (4.2,3.4); 
\end{tikzpicture}
}
\put(37,78){\textcolor{white}{\bf SCoTSS}}
\put(130,0)
{
\begin{tikzpicture}
    \node[anchor=south west,inner sep=0] at (0,0) {\includegraphics[width=0.47\textwidth]{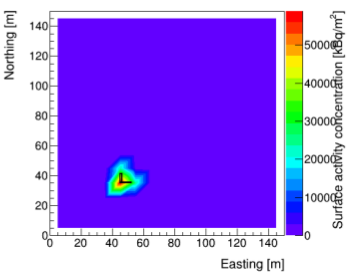}};
    \draw [fill=white,white] (3.64,0.1) rectangle (4.2,3.4); 
\end{tikzpicture}
}
\put(167,78){\textcolor{white}{\bf ARDUO}}\end{picture}
\caption{Response-function fit reconstruction}
\end{subfigure}
\caption{Images of distributed sources made from 150~m $\times$ 150~m
  grid surveys at 10~m altitude. Panels (a), (c) and (e) show
  reconstructions of a 100~m-scale L-shaped distributed source with a
  10~m $\times$ 60~m northing section and a 70~m $\times$ 5~m easting
  section. Panels (b), (d) and (f) show reconstructions of a
  10~m-scale L-shaped distributed source with a 1~m $\times$ 6~m
  northing section and a 7~m $\times$ 0.5~m easting section. In panels
  (c) and (d) the back-projection method is applied to the SCoTSS
  instrument and the simple direction finding method is applied to
  ARDUO. The rate survey, back-projection and simple direction finding
  results are available in real-time, since they don't employ
  computationally intensive simulations or algorithms, unlike the
  response-function fit reconstructions shown in panels (e) and
  (f). Notable is the fact that the self-shielding methods fail to
  reconstruct the spatial detail of the 10~m-scale L-shaped source.}
\label{fig_L}
\end{figure*}

\section{Results}
\subsection{Point-source results}
The results of our point-source comparative studies are shown in
Figures~\ref{fig_tti}a-c, where we plot the reconstruction precision obtained
against data acquisition time. We call this metric the ``time-to-image''
\cite{2014ITNS...61.2745S}. The reconstruction precision is formally the root
mean square (RMS) spread of reconstructed image directions computed from a
sample of independent datasets of a certain acquisition time. Generally, as
the acquisition time increases, greater and greater angular precision is
achieved. The circular symbols in Figures~\ref{fig_tti}d-f show individual
ARDUO self-shielding \textit{response-function search} reconstructions, with
their colour (from light-grey to black) indicating the acquisition time. These
data are used to compute the ARDUO self-shielding RMS values plotted in
Figures~\ref{fig_tti}a-c. The colour maps plotted in Figures 2a-c indicate the
likelihood that the self-shielding algorithm will reconstruct a given position
in 4$\pi$ phase space when the true position is located at an azimuth of
90$^{\circ}$ and a polar angle stated in the individual sub-captions. The
plotted colour indicates the $N$-dimensional distance in LUT rate-space between
two locations in the LUT, as discussed in
Section~\ref{sss1}. Figures~\ref{fig_tti}g-i show, for SCoTSS self-shielding
reconstructions, the same information as Figures~\ref{fig_tti}d-i show for the
ARDUO reconstructions. 

Notable is the likelihood information in Figure~\ref{fig_tti}h, indicating
that the crystal configuration in the SCoTSS instrument yields areas of the
response function LUT which are self-similar, leading to erroneous
reconstructions which do not significantly improve with accumulation
time. This is directly responsible for the large RMS of the SCoTSS
self-shielding reconstruction plotted in Figure~\ref{fig_tti}b. The likelihood
information in Figure~\ref{fig_tti}i also illustrates that for sources which
are directly side on (impinging at a polar angle of 90$^\circ$) the LUT
information is highly non-degenerate, resulting in the relatively small SCoTSS
RMS values plotted in Figure~\ref{fig_tti}c. This wide range of pointing
precision, occurring over a relatively narrow range of incident angles
(30$^\circ$), serves to illustrate a specific finding of these studies - that
the self-shielding performance is highly sensitive to detector configuration
and source location.

\begin{figure}
\centering
\includegraphics[width=0.37\textwidth]{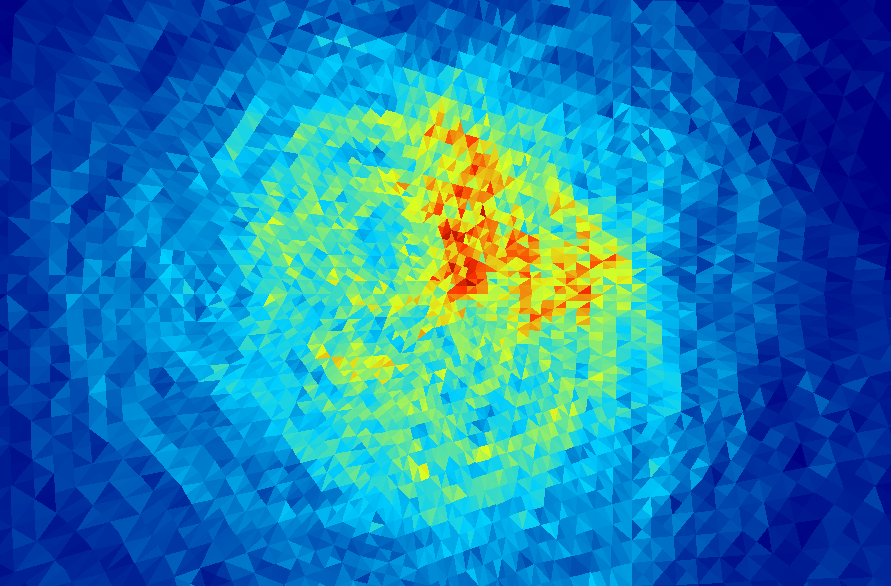}
\caption{Zoomed image of 10~m-sized L-shaped source}
\caption{Back-projection Compton imaging of the 10~m-sized L-shaped
  distributed source with a 1~m $\times$ 6~m vertical section and a 7~m
  $\times$ 0.5~m horizontal section. The colour scale indicate the number of
  overlapping back-projected cones. The lower left vertex of the ``L'' is
  located in the centre of the image which is projected on to a plane at a
  distance of 10~m. The horizontal range is 112$^\circ$ and the vertical range
  is 90$^\circ$. The image corresponds to an acquisition time of 1~second. }
\label{fig_sLC}
\end{figure}

\subsection{Distributed-source results}
Reconstructed images of the 100~m-sized and 10~m-sized sources from both the
ARDUO and SCoTSS instruments are shown in Figure~\ref{fig_L}. Here we provide
a qualitative comparison of the reconstructed methods applied, with
quantitative analyses left for a future publication.

\subsubsection{100~m-sized L-shaped source}
The reconstructed images of the 100~m-sized L-shaped source are shown
in Figures~\ref{fig_L}a, c and e. It is clear from Figure~\ref{fig_L}a
that, using the \textit{rate survey} method, the performance of both
instruments is similar, as would be expected.  It is also clear that,
using this method, a grid survey of this distributed source with a
10~m line spacing can recover some of the spatial information, but
much of the fine detail is absent.  Figure~\ref{fig_L}c indicates how
the reconstruction improves when real-time directional information is
utilized. In the case of SCoTSS, the image is produced using a
back-projection algorithm, with the colour scale indicative of the
number of overlapping cones. The neighbouring ARDUO image is formed
using the \textit{simple direction finding} method, with the colour
scale indicative of the number of back-projected vectors. It is clear
that some finer detail is reconstructed using these methods when
compared to the \textit{rate survey} method, and the performance of
both instruments in this setting is roughly comparable, with the
SCoTSS imager more accurately reproducing the true shape.

Figure~\ref{fig_L}e plots the reconstructions using the self-shielding
\textit{response-function fit} method with both instruments. Such analyses
require computationally intensive simulations and fitting algorithms and are
thus only available in post-processing, after the survey is
completed. Noteworthy, however, is that the fine-detail of the distributed
source is largely recovered and the performance of both instruments appears
comparable.
\subsubsection{10~m-sized L-shaped source}
Reconstruction of a distributed source is challenging for any raster
survey method where the grid spacing is similar in size to, or larger
than, the source dimensions (see the reconstructions plotted in
Figures~\ref{fig_L}b, d and f). It is clear that the 10~m-sized
L-shaped source can be detected and its position is relatively well
determined within the 150~m $\times$ 150~m grid. Little fine detail
can be recovered with the ARDUO or SCoTSS instruments, when using
purely non-imaging directional or survey methods. The finer details of
the source are, however, recoverable with the SCoTSS back-projection
method (see Figure~\ref{fig_sLC}), illustrating the imaging
capabilities which motivated the design of the SCoTSS.\linebreak \newpage \noindent

\section{Concluding remarks and discussion}
We have performed a comparative study of the ARDUO and SCoTSS
instruments, which are quite similar in terms of their sensor
materials, active volume, and electrical components. It is clear that
both instruments can localize a point source of emission, with SCoTSS
generally faster than ARDUO at achieving a given level of
precision. We note that, in self-shielding mode, the performance of
both instruments is highly sensitive to the source location. This is
particularly true for the SCoTSS imager which, in certain areas of
angular phase space can perform poorly, but in others, may out-perform
the imaging mode in point source localization. Of course, with SCoTSS,
both reconstruction modes can work in tandem and this is a direction
we will pursue in the future.

Both instruments have proven capable of imaging a 100~m-sized
distributed source in a raster survey, with sharper images obtained in
real time when the directional information is included in the
reconstruction. Post-acquisition processing methods, such as
response-function deconvolution, can further improve the results from
both detectors. In a setting such as this - a raster survey with 10~m
grid spacing over a 100~m-sized distributed source - our studies show
that the ARDUO and SCoTSS instruments perform in a very similar way in
head-to-head comparisons in all the methods we tested. We arrive at
similar conclusions from our imaging studies performed with a
10~m-sized distributed source, with the notable exception that the
SCoTSS instrument can resolve features small in dimension in
comparison to the survey parameters, unlike ARDUO. We note, of course,
that there are many other relevant deployment scenarios which have not
been investigated here. In particular, situations where multiple
sources or distributed sources are observed from a stationary
position. Imaging capabilities are required in such settings and thus
the SCoTSS detector will far outperform purely directional
instruments, like ARDUO, in these areas.


\end{document}